\documentclass{article}

\usepackage{arxiv}

\usepackage[utf8]{inputenc} 
\usepackage[T1]{fontenc}    
\usepackage{hyperref}       
\usepackage{url}            
\usepackage{booktabs}       
\usepackage{amsfonts}       
\usepackage{nicefrac}       

\usepackage{amssymb}
\usepackage{latexsym}

\usepackage{amsmath}
\usepackage{dsfont}

\usepackage{pgfplotstable}
\usepackage[makeroom]{cancel}
\pgfplotsset{compat=newest}
\usepackage{booktabs}

\usepackage{algorithm}
\usepackage{algorithmic} 

\usepackage{glossaries}
\newacronym{fem}{FEM}{finite element method}
\newacronym{fe}{FE}{finite element}
\newacronym{mc}{MC}{Monte Carlo}
\newacronym{qoi}{QoI}{quantity of interest}
\newacronym{pec}{PEC}{perfect electric conductor}

\usepackage[utf8]{inputenc}

\usepackage{mathtools}
\mathtoolsset{showonlyrefs}

\usepackage{comment}

\usepackage{tikz}
\usepackage{circuitikz}
\usepackage{pgfplots}
\usetikzlibrary{arrows,calc,shapes,arrows,positioning,shadows.blur,shapes.symbols,patterns,decorations.markings, intersections, fillbetween}
\usetikzlibrary{matrix}

\usepackage{graphicx}

\newcommand{\EE}{\mathbb{E}}

\newcommand{\RR}{\mathbb{R}}

\newcommand{\One}{\mathbf{1}}
\newcommand{\intd}{\, \text{d}}
\newcommand{\vm}[1]{\ensuremath{\mathbf{#1}}}
\newcommand{\vms}[1]{\ensuremath{\boldsymbol{#1}}}

\newcommand{\transpose}{\text{T}}

\newcommand{\pdf}{\text{pdf}}
\newcommand{\up}{\vm p} 
\newcommand{\dep}{\vm d} 
\newcommand{\rp}{r}  
\newcommand{\Nmc}{N_{\text{MC}}} 
\newcommand{\QoI}{Q} 
\newcommand{\SDd}{\Omega_{\text{\dep}}} 
\newcommand{\IRP}{T_{\rp}} 

\title{Efficient yield optimization with limited gradient information}

\date{}

\renewcommand{\headeright}{}
\renewcommand{\undertitle}{}

\begin{document}
\maketitle

\vspace*{-2cm}

\begin{center}
{\large Mona Fuhrländer$^{1,2}$ and Sebastian Schöps$^{1,2}$}\\
$^1$ Computational Electromagnetics Group (CEM), Technische Universität Darmstadt, Germany\\
$^2$ Centre for Computational Engineering (CCE), Technische Universität Darmstadt, Germany\\
\end{center}

\vspace*{1cm}

\begin{abstract}
In this work an efficient strategy for yield optimization with uncertain and deterministic optimization variables is presented. The gradient based adaptive Newton-Monte Carlo method is modified, such that it can handle variables with (uncertain parameters) and without (deterministic parameters) analytical gradient information. This mixed strategy is numerically compared to derivative free approaches.
\end{abstract}

\section{Introduction}
\label{sec:intro}

In mass production one often has to deal with uncertainties due to manufacturing imperfections, which lead to deviations in the specified design parameters, i.e., geometry or material parameters, of the manufactured device. These deviations in the design parameters, may lead to deviations in the performance quantities, such that predefined performance requirements are not fulfilled. Thus, the device is useless. This is of course a waste of time, money and resources -- and should be avoided. 

In order to quantify the uncertainty, we consider the yield as the so-called \textit{probability of success}. It is defined as the percentage of realizations in a manufacturing process, which fulfills all performance requirements, taking into account manufacturing uncertainties~\cite{Graeb_2007aa}. 
The yield can be estimated e.g. by a Monte Carlo (MC) analysis~\cite[Chap. 5]{Hammersley_1964aa}. 
In this work we will focus on the optimization procedure, i.e., the maximization of the yield in order to reduce the negative impact of uncertainty. In~\cite{Graeb_2007aa,Fuhrlander_2020aa} gradient based optimization algorithms have been proposed, assuming that gradients are available in analytical form. But this is only the case under some suitable conditions. In the following we present a strategy for efficient yield optimization under the assumption that only some of the partial derivatives are available.

\section{Definition of the yield}
\label{sec:def}

We define three kinds of parameters: uncertain design parameters, deterministic design parameters and range parameters. The uncertain parameters $\up$ are modeled as normal distributed random variables, i.e.,
\begin{equation}
\up \sim {\mathcal{N}\left( \overline{\up}, \vms{\Sigma}\right)}, \text{ with } 
\pdf_{\mathcal{N}\left( \overline{\up}, \vms{\Sigma}\right)}(\up) = \exp\left({-\frac{1}{2}(\up-\overline{\up})^{\transpose} \vms \Sigma^{-1} (\up-\overline{\up})}\right),
\label{eq:UQPara}
\end{equation}
where $\overline{\up} \in \RR^{n_{\up}}$ indicates the mean value, $\vms{\Sigma}\in \RR^{n_{\up} \times n_{\up}}$ the covariance matrix and $\pdf_{\mathcal{N}\left( \overline{\up}, \vms{\Sigma}\right)}$ the corresponding probability density function (pdf).
The deterministic parameters are given by $\dep \in \RR^{n_{\dep}}$. The range parameter is denoted by $\rp \in \IRP \subset \RR$ and describes the environment in which the requirements have to be fulfilled.

Let $\QoI:\RR^{n_{\up}+n_{\dep}+1} \rightarrow \RR$ be a quantity of interest (QoI) and $c \in \RR$. We define the performance feature specifications (pfs) as
\begin{equation}
\QoI_r(\up,\dep) \leq c \quad \forall r\in \IRP,
\label{eq:QoI}
\end{equation}
which can be easily extended to a vector-valued formulation in case of several requirements.
Then the safe domain is the set of all parameter combinations fulfilling the pfs, and it depends on the current value of $\dep$, i.e.,
\begin{equation}
\SDd = \left\lbrace  \up: \QoI_r(\up,\dep) \leq c \ \ \forall \rp \in  \IRP \right\rbrace.
\label{eq:SafeDomain}
\end{equation}
The yield $Y$ defines the percentage of realizations in a manufacturing process, which fulfill the pfs. Following~\cite{Graeb_2007aa} it is given by
\begin{equation}
Y(\overline{\up},\dep) := \EE [\One_{\SDd}(\up)]
:=\int_{\RR^{n_{{\scalebox{.35}{\up}}}}} \One_{\SDd}(\up) \, \pdf_{\mathcal{N}\left( \overline{\up}, \vms{\Sigma}\right)}(\up)  \intd \up,
\label{eq:Yield}
\end{equation}
where $\EE$ denotes the expected value and $\One_{\SDd}$ the indicator function with value $1$ if the parameter $\up$ lies inside the safe domain and $0$ otherwise.

A straightforward approach for yield estimation is MC analysis~\cite[Chap. 5]{Hammersley_1964aa}. There, a large set of sample points $\up^{(1)},\dots,\up^{(\Nmc)}$ of the uncertain parameter is randomly generated according to its pdf. Then the yield can be estimated by
\begin{equation}
Y(\overline{\up},\dep) \approx Y_{\text{MC}}(\overline{\up},\dep) 
= \frac{1}{\Nmc} \sum_{i=1}^{\Nmc} \One_{\SDd}(\up^{(i)}).
\label{eq:YieldMC}
\end{equation}
In computational engineering, the QoI often involves solving partial differential equations numerically, e.g., with a finite element method (FEM). Hence, it is computationally very expensive or even prohibitive to evaluate the QoI for the many sample points required in a MC analysis. For that reason, there is research on efficient yield estimation, using e.g. importance sampling~\cite{Gallimard_2019aa}, surrogate modeling~\cite{Babuska_2007aa,Rasmussen_2006aa,Rao_1999aa} or hybrid approaches~\cite{Li_2010aa,Fuhrlander_2020aa,Fuhrlander_2020ab}. These hybrid approaches combine classic MC with surrogate methods, e.g. Gaussian process regression (GPR), cf.~\cite{Fuhrlander_2020ab}. Since this work focuses on the optimization process, we will not go into the details here. 

\section{Yield optimization}
\label{sec:opt}
We aim to maximize the yield by modifying the design, i.e.,
\begin{equation}
\max_{\overline{\up},\dep} Y(\overline{\up},\dep).
\label{eq:maxYield}
\end{equation}
Let us assume that we have only uncertain design parameters as optimization variables and all of them are Gaussian distributed, i.e., $\max_{\overline{\up}} Y(\overline{\up},\dep)$. Then there exist closed form solutions of gradient and Hessian, cf.~\cite{Graeb_2007aa},
\begin{align}
\nabla_{\overline{\up}} Y(\overline{\up},\dep) =  \int_{\RR^{n_{{\scalebox{.35}{\up}}}}} \One_{\SDd}(\up) \, \nabla_{\overline{\up}}\pdf_{\mathcal{N}\left( \overline{\up}, \vms{\Sigma}\right)}(\up)  \intd \up,
\label{eq:GradYield} \\
\nabla^2_{\overline{\up}} Y(\overline{\up},\dep) = \int_{\RR^{n_{{\scalebox{.35}{\up}}}}} \One_{\SDd}(\up) \, \nabla^2_{\overline{\up}}\pdf_{\mathcal{N}\left( \overline{\up}, \vms{\Sigma}\right)}(\up)  \intd \up,
\label{eq:HessYield}
\end{align}
since the optimization variable $\overline{\up}$ only appears in the pdf and this is just an exponential function in case of Gaussian distribution.
The MC estimators of the gradient and the Hessian are given by
\begin{align}
\nabla_{\overline{\up}} Y_{\text{MC}}(\overline{\up},\dep) &= 
Y_{\text{MC}}(\overline{\up},\dep) \Sigma^{-1} \left(\overline{\up}_{\SDd} - \overline{\up}\right),
\label{eq:MCGradYield} \\
\nabla^2_{\overline{\up}} Y_{\text{MC}}(\overline{\up},\dep) &= 
Y_{\text{MC}}(\overline{\up},\dep) \Sigma^{-1} 
\left(\vms \Sigma_{\SDd} + \left(\overline{\up}_{\SDd} - \overline{\up}\right)
\left(\overline{\up}_{\SDd} - \overline{\up}\right)^{\transpose} - \vms \Sigma  \right) \Sigma^{-1},
\label{eq:MCHessYield}
\end{align}
where $\overline{\up}_{\SDd}$ indicates the mean value of all MC sample points lying inside the safe domain and $\vms \Sigma_{\SDd}$ the corresponding covariance matrix. The detailed derivation can be found in~\cite{Graeb_2007aa}. Using~\eqref{eq:MCGradYield} and~\eqref{eq:MCHessYield}, once the yield is estimated with MC, the derivatives are obtained without any additional computational effort. This allows to use a gradient based optimization solver, e.g. a globalized Newton method, cf.~\cite{Ulbrich_2012aa}. 

In~\cite{Fuhrlander_2020aa} an adaptive Newton-MC method is proposed, which is an efficient modification of the globalized Newton method using the standard deviation of the MC estimation 
\begin{equation}
\sigma_{\text{MC}}(\overline{\up},\dep) = \sqrt{\frac{Y_{\text{MC}}(\overline{\up},\dep)(1-Y_{\text{MC}}(\overline{\up},\dep))}{\Nmc}}
\end{equation}
as an error indicator for an adaptive sample size increase. For details we refer to~\cite{Fuhrlander_2020aa}.

Back to problem~\eqref{eq:maxYield}, we have uncertain \textit{and} deterministic optimization variables. Since $\dep$ appears in the indicator function, we cannot calculate the gradient of the yield with respect to $\dep$, given by
\begin{equation}
\nabla_{\dep} Y(\overline{\up},\dep) 
=  \int_{\RR^{n_{{\scalebox{.35}{\up}}}}} \nabla_{\dep} \One_{\Omega_{\dep}}(\up) \, \pdf_{\mathcal{N}\left( \overline{\up}, \vms{\Sigma}\right)}(\up)  \intd \up,
\label{eq:GradYield_d}
\end{equation}
analytically. Same holds for the Hessian. In order to still use the globalized Newton method or the adaptive Newton-MC, we propose a mixed strategy. We calculate the gradient with respect to $\dep$ with finite differences. But we still use the analytical form for the derivative with respect to $\up$. So we have 
\begin{equation}
\nabla_{\overline{\up},\dep} Y(\overline{\up},\dep) = \left( \nabla_{\overline{\up}} Y(\overline{\up},\dep), \nabla_{\dep} Y(\overline{\up},\dep) \right)^{\transpose},
\label{eq:GradCompl}
\end{equation}
where the first part is calculated with~\eqref{eq:GradYield} and the second part with finite differences.
A well-known formula to approximate Hessians is the Broyden-Fletcher-Goldfarb-Shanno (BFGS) update~\cite{Ulbrich_2012aa}, given by
\begin{equation}
\vm H_{k+1}^{\text{BFGS}} = 
\vm H_k + \frac{\vm g_k \vm g_k^{\transpose}}{\vm g_k^{\transpose} \vm x_k}
- \frac{\vm H_k \vm x_k (\vm H_k \vm x_k)^{\transpose}}{\vm x_k^{\transpose} \vm H_k \vm x_k},
\label{eq:HessBFGS}
\end{equation}
where $\vm H_k$ is the Hessian from the last iterate, $\vm g_k$ the difference between the current and the last gradient and $\vm x_k$ the difference between the current and the last solution.
Since the part of the Hessian belonging to the uncertain parameter can be calculated analytically by~\eqref{eq:HessYield}, we introduce the mixed BFGS Hessian
\begin{equation}
\vm H_{\text{mix}}^{\text{BFGS}} := 
\begin{pmatrix}
\nabla_{\overline{\up}}^2 Y(\overline{\up},\dep) & \vline & \vm H^{\text{BFGS}} \\
\hline
\vm H^{\text{BFGS}}  & \vline & \vm H^{\text{BFGS}}
\end{pmatrix}
\in \RR^{\left(n_{\up}+n_{\dep}\right) \times \left(n_{\up}+n_{\dep}\right)},
\label{eq:HessMix}
\end{equation}
where we insert the analytical Hessian $\nabla_{\overline{\up}}^2 Y(\overline{\up},\dep) \in \RR^{n_{\up}\times n_{\up}}$ from~\eqref{eq:HessYield} into the BFGS formulation~\eqref{eq:HessBFGS}.
The mixed strategy can also be necessary, if the gradient or Hessian of the pdf cannot be computed in closed form.

\section{Numerical results}
\label{sec:num}

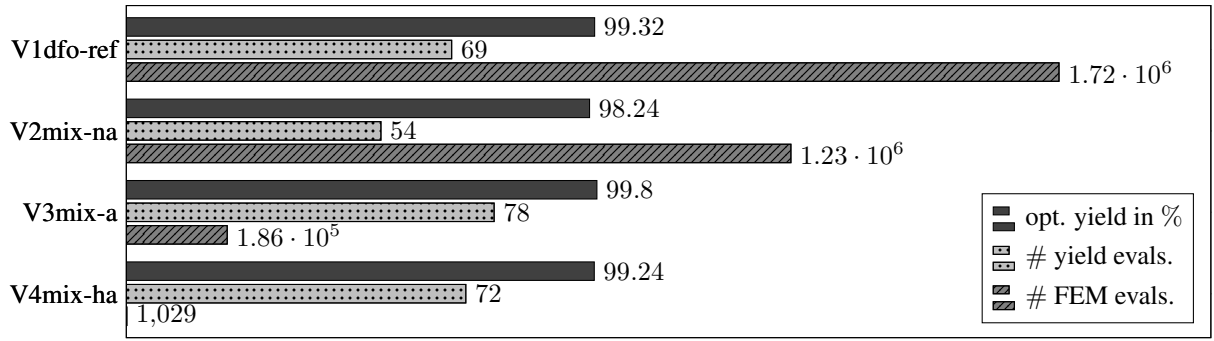
\begin{figure}[t]
	\begin{tikzpicture}[scale=1.]
	\pgfplotsset{
		height=6cm, width=16cm,
		every axis/.style={
			symbolic y coords = {V4mix-ha,V3mix-a,V2mix-na,V1dfo-ref}}
	}
	\begin{axis}[
	xbar,
	reverse legend,
	legend pos=south east,
	legend cell align=left,
	/pgf/bar width=7pt,
	tickwidth         = 0pt,
	enlarge y limits  = 0.18,
	ytick             = data,
	xmin=0,xmax=230,
	xtick = \empty,
	nodes near coords,
	]
	\addplot[fill=lightgray,postaction={pattern=dots},bar shift=0.0cm] coordinates { (72,V4mix-ha)(78,V3mix-a)(54,V2mix-na)(69,V1dfo-ref)};\label{pgf:YE} 
	\addplot[fill=darkgray,bar shift=0.3cm] coordinates { (99.24,V4mix-ha)(99.80,V3mix-a)(98.24,V2mix-na)(99.32,V1dfo-ref)};\label{pgf:Yield}
	\end{axis}
	\begin{axis}[
	xbar,
	reverse legend,
	legend pos=south east,
	legend cell align=left,
	/pgf/bar width=7pt,
	tickwidth         = 0pt,
	enlarge y limits  = 0.18,
	ytick             = data,
	xmin=0,xmax=2000000,
	xtick = \empty,
	nodes near coords,
	]
	\addplot[fill=gray,postaction={pattern=north east lines},bar shift=-0.3cm] coordinates { (1029,V4mix-ha)(186070,V3mix-a)(1225592,V2mix-na)(1719976,V1dfo-ref)};\label{pgf:HF}
	\end{axis}
	%
	%
	\matrix[
	matrix of nodes,
	anchor=north west,
	draw,
	inner sep=0.2em,
	column 1/.style={nodes={anchor=center}},
	column 2/.style={nodes={anchor=west},font=\strut},
	draw
	]
	at([xshift=-3.05cm,yshift=-2.5cm]current axis.north east){
		\ref{pgf:Yield}& opt. yield in $\%$\\
		\ref{pgf:YE}& $\#$ yield evals.\\
		\ref{pgf:HF}& $\#$ FEM evals.\\}	
	;
	\end{tikzpicture}
	\caption{Comparison of different methods for yield optimization.}
	\label{fig:Comp4}
\end{figure}

As benchmark problem we consider a simple dielectrical waveguide with two uncertain geometrical parameters $p_1$ (length of the inlay) and $p_2$ (length of the offset) and two deterministic material parameters $d_1$ and $d_2$ with impact on the relative permittivity and permeability. The uncertain parameters are assumed to be independent truncated Gaussian distributed with truncation at $\pm3\,\text{mm}$ in order to avoid unphysical values. Thus, the parameters and their initial values for optimization are given by
\begin{equation}
\overline{\up}^0 = \left[9,5\right], \ 
\vms \Sigma = \text{diag}\left(\left[0.9^2,0.9^2\right]\right) \text{ and } 
\dep^0 = \left[1,1\right].
\end{equation}
The range parameter is the angular frequency.
The QoI is the scattering parameter (S-parameter), i.e., for its calculation the electric field formulation of Maxwell has to be solved numerically with FEM. We consider the pfs
\begin{equation}
\QoI_r(\up) \leq -24\,\text{dB} \quad \forall r\in \IRP=\left[2\pi 6.5,2\pi 7.5\right] \text{ in GHz.}
\label{eq:QoIwg}
\end{equation}
The frequency range $\IRP$ is discretized into $11$ equidistant frequency points. For each of these points, the inequality in~\eqref{eq:QoIwg} has to be fulfilled. For more details regarding this example we refer to~\cite{Loukrezis_WG} and~\cite{Fuhrlander_2020aa}.
In the optimization we set $\sigma_{\text{MC}}^{\text{max}} = 0.01$, which implies $\Nmc = 2500$ in the non-adaptive method. In the adaptive Newton-MC we set $\Nmc^0 = 100$ and increase it if necessary.
The initial yield value is $Y_{\text{MC}}^0 = 42.8\,\%$. 
We compare four methods to maximize the yield of this waveguide:
\begin{itemize}
	\item V1dfo-ref: reference solution -- problem solved with classic MC for estimation and the derivative free optimization (DFO) solver Py-BOBYQA~\cite{Cartis_2019aa}
	\item V2mix-na: mixed strategy proposed in Section~\ref{sec:opt} with classic MC for estimation and non-adaptive Newton method for optimization
	\item V3mix-a: mixed strategy proposed in Section~\ref{sec:opt} with classic MC for estimation and adaptive Newton-MC for optimization
	\item V4mix-ha: mixed strategy proposed in Section~\ref{sec:opt} with Hybrid-GPR approach~\cite{Fuhrlander_2020ab} for estimation and adaptive Newton-MC for optimization
\end{itemize}
We consider three aspects of these methods: the optimal yield they achieve, the number of objective function (i.e. yield) calls they require and the number of FEM evaluations (to solve the QoI). The results are shown in Figure~\ref{fig:Comp4}.

All methods achieve an improvement of the yield by more than $55\,\%$ to values between $98.4\,\%$ (V2mix-na) and $99.8\,\%$ (V3mix-a), with the optimal solutions
\begin{align*}
&\text{V1dfo-ref:} && \hspace{-1cm}\overline{\up}^{\text{opt}} = \left[10.94,5.22\right] \text{ and } \dep^{\text{opt}} = \left[0.44,1.19\right]\\
&\text{V2mix-na:} && \hspace{-1cm}\overline{\up}^{\text{opt}} = \left[10.56,4.52\right] \text{ and } \dep^{\text{opt}} = \left[0.39,1.16\right]\\
&\text{V3mix-a:} && \hspace{-1cm}\overline{\up}^{\text{opt}} = \left[\ \ 9.87,4.92\right] \text{ and } \dep^{\text{opt}} = \left[0.2,0.125\right]\\
&\text{V4mix-ha:} && \hspace{-1cm}\overline{\up}^{\text{opt}} = \left[10.86,5.22\right] \text{ and } \dep^{\text{opt}} = \left[0.44,1.09\right]
\end{align*}
In the non-adaptive case (V1dfo-ref and V2mix-na) the number of yield evaluations correlates strongly with the number of FEM evaluations. The mixed strategy from Section~\ref{sec:opt} (V2mix-na) needs more than $20\,\%$ less yield and FEM evaluations than the reference DFO solver Py-BOBYQA (V1dfo-ref). When introducing the adaptive Newton-MC (V3mix-a and V4mix-ha), the number of yield evaluations increases, which can be explained by less accurate descent directions due to noisier yield estimations because of smaller MC sample sets. Nevertheless, the total computational effort, i.e., the number of FEM evaluations, decreases, since the yield evaluations are run with smaller MC sample sets and are thus less expensive. By not applying classic MC for yield estimation, but a hybrid approach based on GPR surrogates (V4mix-ha), the computational effort again can be reduced by a factor of $180$ compared to classic MC (V3mix-a), by $1191$ compared to the non-adaptive strategy (V2mix-na) and by $1671$ compared to the DFO reference (V1dfo-ref).

\section{Conclusion}
\label{sec:concl}

We proposed a new mixed approach to solve yield optimization problems with deterministic and uncertain optimization variables. Only for the uncertain parameters, analytical gradient and Hessian information is available. Thus, a mixed strategy with analytical and numerical (finite differences and BFGS updates) derivatives has been used. Numerical results show better efficiency than a common derivative free optimization solver. Future research will deal with implementing the adaptive strategy and available gradient information into an originally derivative free solver.

\section*{Acknowledgements}
The work of Mona Fuhrländer is supported by the Graduate School CE within the Centre for Computational Engineering at Technische Universität Darmstadt.

\end{document}